\documentclass[journal]{IEEEtran}
\usepackage[latin9]{inputenc}
\usepackage{graphicx}

\usepackage[cmex10]{amsmath}
\usepackage{amssymb}
\usepackage{array}

\usepackage{eqparbox}

\usepackage{srcltx}
\usepackage{fixltx2e}

\usepackage{url}

\begin{document}

\title{Digital Holography at Shot Noise Level}

\author{Frédéric~Verpillat,
        Fadwa~Joud,
        Michael~Atlan
        and~Michel~Gross
\thanks{Frédéric~Verpillat, Fadwa~Joud and Michel~Gross are with the
 Laboratoire Kastler Brossel. \'Ecole Normale Sup\'erieure ,
  UMR 8552 , UPMC, CNRS
24 rue Lhomond , 75231 Paris Cedex 05; France  e-mail: frederic.verpillat@lkb.ens.fr; fadwa.joud@lkb.ens.fr and michel.gross@lkb.ens.fr}
\thanks{Michael~Atlan is with the Fondation Pierre-Gilles de Gennes \& Institut Langevin: UMR 7587 CNRS INSERM, ESPCI ParisTech, Universit\'e Paris 6, Universit\'e Paris 7, 10
rue Vauquelin, 75 231 Paris Cedex 05, France. e-mail: atlan@optique.espci.fr}
\thanks{Manuscript received \today}}

\markboth{IEEE~/~OSA Journal of Display Technology}%
{Shell \MakeLowercase{\textit{et al.}}: Bare Demo of IEEEtran.cls for Journals}


\maketitle

\begin{abstract}
By a proper arrangement of a digital holography setup, that combines off-axis  geometry with phase-shifting recording conditions, it is possible to reach the theoretical shot noise limit, in real-time experiments. We studied this limit, and we show that it corresponds to 1 photo-electron per pixel within the whole frame sequence that is used to reconstruct the holographic image. We also show that Monte Carlo noise synthesis onto holograms measured at high illumination levels enables accurate representation of the experimental holograms measured at very weak illumination levels. An experimental validation of these results is done.

\end{abstract}

%

%
\IEEEpeerreviewmaketitle

\section{Introduction}

%
%

\IEEEPARstart{D}{emonstrated} by Gabor \cite{Gabor49} in the early 50's, the purpose of holography is to record,
on a 2D detector, the phase and the amplitude of the radiation field scattered from an object under coherent illumination. The photographic film used in conventional holography is replaced by a 2D electronic detection in digital holography \cite{Macovsky1971}, enabling quantitative numerical analysis. Digital holography has been waiting for the recent development of computer and video technology to be experimentally demonstrated
\cite{Schnars94}. The main advantage of digital holography is that, contrary to holography with photographic plates
\cite{Gabor49}, the holograms are recorded by a photodetector array, such as a CCD camera, and the image is digitally reconstructed by a computer, avoiding photographic processing \cite{Goodmann_1967}.

Off-axis holography \cite{leith1965microscopy} is the oldest configuration adapted to digital
holography \cite{Schnars_Juptner_94,Schnars94,Kreis88}. In off-axis digital holography, as well as in photographic
plate holography, the reference beam is angularly tilted with respect to the object observation axis. It is then possible to record, with a single hologram, the two quadratures of the object's complex
field. However, the object field of view is reduced, since one must avoid the overlapping of the image with the
conjugate image alias \cite{Cuche00}. Phase-shifting digital holography, which has been introduced later
\cite{Yamaguchi1997}, records several images with a different phase for the reference beam. It is then possible to obtain the
two quadratures of the field in an in-line configuration even though the conjugate image alias and the true image
overlap, because aliases can be removed by taking image differences.

With the development of CCD camera technologies, digital holography became a fast-growing research field that has drawn
increasing attention \cite{schnars2002digital,doval_2000}. Off-axis holography has been applied recently to particle
    \cite{pu2004intrinsic}
    polarization \cite{colomb2002polarization},
    phase contrast \cite{Cuche99},
    synthetic aperture \cite{massig2002digital},
    low-coherence \cite{ansari2001elimination,massatsch2005time}
     photothermal \cite{absil2010photothermal},
  and microscopic \cite{massatsch2005time,marquet2005digital,atlan2008heterodyne}
    imaging.
Phase-shifting holography has been applied to
    3D \cite{zhang1998three,nomura2007polarization},
    color \cite{yamaguchi2002phase},
    synthetic aperture \cite{leclerc2001sae},
   low-coherence \cite{tamano2006phase},
    surface shape \cite{yamaguchi2006surface},
    photothermal \cite{absil2010photothermal},
     and microscopic \cite{zhang1998three,yamaguchi2001image,atlan2008heterodyne} imaging.

We have developed an alternative phase-shifting digital holography technique  that uses a frequency shift of the
reference beam to continuously shift the phase of the  recorded interference pattern \cite{Leclerc2000}. One of the
advantages of our setup is its ability to provide accurate phase shifts    that allow to suppress twin images aliases
\cite{atlan2007accurate}. More generally, our setup can be viewed as a multipixel heterodyne detector that is able of
recording the complex amplitude of the signal electromagnetic field $\cal E$ in all pixels  of the CCD camera in
parallel. We get then the map of the field over the CCD pixels (i.e. ${\cal E}(x,y)$ where $x$ and $y$ are the pixels
coordinate). Since the field is measured on all pixels at the same time, the relative phase that is measured for
different locations $(x,y)$ is meaningful. This means that the field map ${\cal E}(x,y)$ is a hologram that can be used
to reconstruct the field $\cal E$ at any location along the free-space optical propagation axis, in particular in the
object plane.

Our heterodyne holographic setup has been used  to perform holographic \cite{Leclerc2000}, and synthetic aperture
\cite{leclerc2001sae} imaging.  We have also demonstrated that our heterodyne technique used in an off-axis holographic
configuration is capable of recording holograms with optimal sensitivity \cite{gross2007dhu}. This means that it is
possible to fully filter-off technical noise sources, that are related to the reference beam (i.e. to the zero order
image \cite{cuche2000spatial}), reaching thus, without any experimental effort, the quantum limit of noise of one photo
electron per reconstructed pixel during the whole measurement time.

In the present paper we will discuss on noise in digital holography, and we  will try to determine what is the ultimate
noise limit both theoretically, and in actual holographic experiments in real-time. We will see that, in the
theoretical ideal case, the limiting noise is the Shot Noise on the holographic reference beam. In reference to
heterodyne detection, we also refer to the reference beam as Local Oscillator (LO). We will see that the ultimate
theoretical limiting noise can be reached in real time holographic experiments, by combining the two families of
digital holography setups i.e. phase-shifting and off-axis. This combination enables to fully filter-off technical
noises, mainly due to LO beam fluctuations in low-light conditions, opening the way to holography with ultimate
sensitivity \cite{gross2007dhu,gross2008noise}.

\section{Off-axis + Phase-Shifting holography}

In order to discuss on noise limits in digital holography, we first need to give some general information on holography
principles. We will thus describe here a typical digital holographic setup, how the holographic information is obtained
from recorded CCD images, and how this information is used to reconstruct holographic images in different
reconstruction planes. We will consider here the case of an off-axis + phase-shifting holographic setup, able to reach
the ultimate noise limit, in low-light imaging conditions, in real time.

\subsection{The Off-axis + Phase Shifting holography setup}

\begin{figure}[h!]
\begin{center}
\includegraphics[width=8cm,keepaspectratio=true]{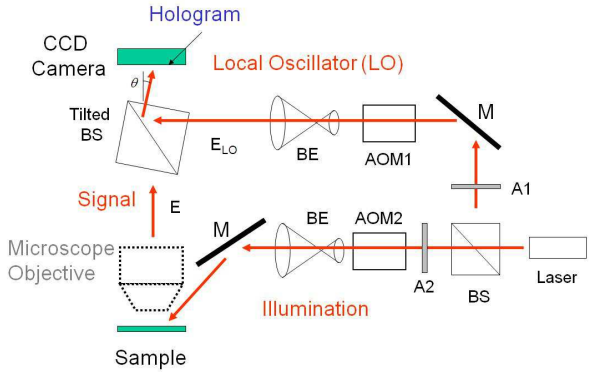}
\caption{Digital holography setup. AOM1 and AOM2 : acousto-optic modulators; BS :  beam splitter; BE : beam expander; M
: mirror; A1 and A2 : attenuator; $\theta$ :  tilt angle of the beam splitter with respect to optical
axis}\label{fig1_setup1}
\end{center}
\end{figure}

The holographic setup used in the following
discussion, is presented on Fig.\ref{fig1_setup1}. We have
considered here, a reflection configuration, but the
discussion will be the same in case of transmission configuration.

The main optical beam (complex field $E_L$, optical angular frequency $\omega_L$) is provided by a Sanyo (DL-7147-201)
diode laser ($\lambda=658\;\rm{nm}$). It is split through a 50/50 Beam Splitter (BS) into an illumination beam
($E_I$,$\omega_I$), and a LO beam ($E_{LO}$,$\omega_{LO}$). The illumination intensity can be reduced with grey neutral
filters. Both beams go through Acousto-Optic Modulators (AOMs) (Crystal Technology, $\omega_{\rm{aom1,2}}\simeq
80\;\rm{MHz}$) and only the first diffraction order is kept. In the typical experiment  case considered here, the
modulators are adjusted for the 4-phase heterodyne detection, but other configurations are possible (8-phases, sideband
detection ...). We have thus :
%
%
\begin{equation}
\omega_I=\omega_L+\omega_{\rm{aom2}}
\end{equation}
\begin{equation}
\omega_{LO}=\omega_L+\omega_{\rm{aom1}}
\end{equation}
with :
\begin{equation}\label{Condition_heterodyne}
\omega_I-\omega_{LO} = 2\pi f_{\rm{ccd}} /4
\end{equation}
where $f_{\rm{ccd}}$ is the acquisition frame rate of the CCD (typically $12.5\;\rm{Hz}$).

The  beams outgoing from the AOMs are expanded by Beam Expanders BEs. The illumination beam  is pointed towards the
object studied. The reflected radiation ($E$,$\omega = \omega_I$) and the LO beam are combined with a beam splitter,
which is angularly tilted by typically $1^{\circ}$, in order to be in an Off-Axis holographic configuration. Light can
be collected with an objective for microscopic imaging. Interferences between reflected light and LO are recorded with
a digital camera (PCO Pixelfly): $f_{\rm{ccd}}=12.5\;\rm{Hz}$, $1280\times1024$ pixels of $6.7\times6.7\;\rm{\mu m}$,
12-bit.

We can notice that our Off-axis + Phase-Shifting (OPS) holographic setup, presented here, exhibits several advantages.
Since we use AOMs, the amplitude, phase and  frequency  of both illumination and LO beams can be fully controlled. The
phase errors in phase-shifting  holography can thus be highly reduced \cite{atlan2007accurate}. By playing with the LO
beam frequency , it is possible to get holographic images at sideband frequencies of a vibrating object
\cite{joud2009imaging,joud2009fringe},
or to get Laser Doppler images of a flow \cite{atlan2006laser,atlan2006laser}, and image by the way blood flow, in vivo
\cite{atlan2006frequency,atlan2007cortical,atlan2008high}. The OPS holographic setup can also be used as a multi pixel
heterodyne detector able to detect, with a quite large optical étendue (product of a beam solid angular divergence by
the beam area) the light scattered by a sample, and to analyze its frequency spectrum
\cite{gross2005heterodyne,lesaffre2006effect}. This detector can be used to detect photons that are frequency shifted
by an ultrasonic wave \cite{Gross_03,atlan2005pulsed} in order to perform Ultrasound-modulated optical tomography
\cite{wang1997ultrasound,ramaz_opex_2004,wang1998frequency,gross2005theoretical,sui2005imaging,lesaffre2007situ}.

The OPS setup benefits of another major advantage.    By recording several holograms with different phases (since we do
phase shifting), we perform heterodyne detection. We benefit thus on heterodyne gain. Moreover, since the heterodyne
detector is multi pixels, it is possible to combine information on different pixels in order to extract the pertinent
information on the object under study, while removing the unwanted technical noise of the LO beam. As we will show,
because  the setup is off-axis, the object pertinent information can be isolated from  the LO beam noise.  By this way,
we can easily reach, in a real life holographic experiment, the theoretical noise limit, which is related to the Shot
Noise of LO beam.

\subsection{Four phases detection}

In order to resolve the object field information in quadrature  in the CCD camera plane, we will consider, to simplify
the discussion,  the case of 4 phases holographic detection, which is commonly used in Phase Shifting digital
holography \cite{Yamaguchi1997}.

Sequence of $4n$ frames $I_0$ to $I_{4n-1}$ are recorded at $12.5\;\rm{Hz}$. For each frame $I_k$, the signal on  each
pixel $I_{k,p,q}$ (where $k$ is the frame index, and $p,q$ the pixel indexes along the $x$ and $y $ directions) is
measured in Digital Count (DC) units between 0 and 4095 (since our camera is 12-bit). The $1280\times1024$ matrix of
pixels is truncated to a $1024\times1024$ matrix for easier discrete Fourier calculations. For each frame $k$ the
optical signal is integrated over the acquisition time $T=1/f_{\rm{ccd}}$. The pixel signal $I_{k,p,q}$ is thus defined
by :
\begin{eqnarray}\label{Eq_slide_1}
 I_{k,p,q}  = \int_{t_k-T/2}^{t_k+T/2} dt \iint_{(p,q)}{dx}{dy}
\left|E(x,y,t) +E_{LO}(x,y,t)\right|^2
\end{eqnarray}
where  $\iint_{(p,q)} $ represents  the integral  over the pixel $(p,q)$ area, and $t_k$ is the recording time of frame
$k$. Introducing the complex representations ${\cal E}$ and ${\cal E}_{LO}$ of the fields $E$ and $E_{LO}$, we get :
\begin{equation}
E(x,y,t)={\cal E}(x,y)e^{j\omega_I t}+c.c.
\end{equation}
\begin{equation}
E_{LO}(x,y,t)={\cal E}_{LO}(x,y)e^{j\omega_{LO} t}+c.c
\end{equation}
\begin{eqnarray}\label{Eq_slide_2}
 I_{k,p,q} &&=a^2T  \\
 \nonumber &&\left(|{\cal E}_{p,q}|^2+{|\cal E}_{LO}|^2
 +{\cal E}_{p,q} {\cal E}_{LO}^* {\rm e}^{ j \left(\omega_I-\omega_{LO}\right)t_k}+c.c.\right)
\end{eqnarray}
where $a$ is the pixel size. To simplify the notations in
Eq.\ref{Eq_slide_2}, we have considered that the LO field ${\cal
E}_{LO}$ is the same in all locations $(x,y)$, and that signal
field ${\cal E}_{p,q}$  does not vary within the pixel $(p,q)$.
If ${\cal E}_{LO}$ varies with location, one has to replace
${\cal E}_{LO}$ by ${\cal E}_{LO,p,q}$ in Eq.\ref{Eq_slide_2}.

The condition given in Eq.\ref{Condition_heterodyne}   imposed a phase shift of the LO beam  equal to $\pi/2$ from one
frame to the next. Because of this shift, the complex hologram $H$ is obtained by summing the sequence of $4n$ frames
$I_0$ to $I_{4n-1}$ with the appropriate phase coefficient :

\begin{equation}\label{Sum_heterodyne}
H=\sum^{4n-1}_{k=0} (j)^{k} I_k
\end{equation}
where $H$ is a matrix of pixels $H_{p,q}$. We get from Eq.\ref{Eq_slide_2} :
\begin{equation}\label{Sum_heterodyne_2}
H_{p,q}= \sum^{4n-1}_{k=0} (j)^{k} I_{k,p,q} = 4n a^2 T {\cal E}_{p,q} {\cal E}_{LO}^*
\end{equation}
The complex hologram $H_{p,q}$ is thus proportional to the object field ${\cal E}_{p,q}$ with a proportionality factor
that involves the LO field amplitude ${\cal E}_{LO}^*$.

%

\subsection{Holographic Reconstruction of the Image of the Object}

Many numerical methods can be used to reconstruct  the image of
the object. The most common is the convolution method that
involves a single discrete Fourier Transform \cite{Schnars_Juptner_94}.
Here, we will use the angular spectrum method, which involves
two Fourier transforms
\cite{Leclerc2000,leclerc2001sae,yu2005wavelength}. We have made
this choice because this method keeps constant the pixel size in
the calculation of the grid pixel size, which remains ever equal to the CCD
pixel. It becomes then easier to discuss on noise, and noise
density per unit of area.

The hologram $H$ calculated in Eq.\ref{Sum_heterodyne}  is the
hologram in the CCD plane ($z=0$). Knowing the  complex hologram
$H(x,y,0)$ in the CCD plane, the hologram  $H(x,y,z)$ in
other planes ($z\neq 0$) is calculated by propagating  the
reciprocal space hologram $\tilde{H}(k_x,k_y)$, which is
obtained with a fast Fourier transform (FFT), from $z=0$ to $z$.
\begin{equation}\label{eq_FFT1}
\tilde{H}(k_x,k_y,0) = {\rm FFT} \left[H(x,y,0) \right]
\end{equation}
To clarify the notation, we have replaced  here  $H_{p,q}$ by
$H(x,y,0)$ where $x$ and $,y$ represent the coordinates of the
pixel $(p,q)$. By this way, the coordinates of the reciprocal
space hologram $\tilde{H}$ are simply $k_x$ and $k_y$. In the
reciprocal space, the hologram $\tilde{H}$ can be propagated
very simply:
\begin{equation}\label{eq_mult_phase_matrix}
\tilde{H}(k_x,k_y,z)=\tilde{H}(k_x,k_y,0)\tilde{K}(k_x,k_y,z)
\end{equation}
where $\tilde{K}(k_x,k_y,z)$ is a  phase matrix that describes the propagation from 0 to $z$:
\begin{equation}
\tilde{K}(k_x,k_y,z)= \exp\left( \frac{j\lambda z(k_x^2+k_y^2)}{2\pi} \right)
\end{equation}
The reconstructed image $H(x,y,z)$ in $z \ne 0$ is obtained then by reverse Fourier transformation :
\begin{equation}\label{eq_FFT-1}
H(x,y,z)= {\rm FFT}^{-1} \left[\tilde{H}(k_x,k_y,z) \right]
\end{equation}

In the following, we will see that the  major source of noise is
the shot noise on the LO, and we will show that
this noise  corresponds to an equivalent noise  of 1 photon per
pixel and per frame, on the signal beam. This LO noise, which
corresponds to a fully developed speckle,  is essentially
gaussian, each pixel being uncorrelated with the neighbor
pixels. If one considers that the LO beam power is the same for
all pixel locations (which is a very common approximation), the
noise density of this speckle gaussian noise is the same for all
pixels.

In that uniform (or flat-field) LO beam approximation, all the transformations made in the holographic reconstruction
(FFTs: Eq.\ref{eq_FFT1} and Eq.\ref{eq_FFT-1}, and multiplication by a phase matrix: Eq.\ref{eq_mult_phase_matrix}) do
not change the noise distribution, and the noise density. FFTs change a gaussian noise into another gaussian noise,
and, because of the Parceval theorem, the noise density remains the same. The phase matrix multiplication does not
change the noise either, since the phase is fully random from one pixel to the next. Whatever the reconstruction plane,
the gaussian speckle noise on gets in the CCD plane, transforms into another gaussian speckle noise, with the same
noise density.

\section{The Theoretical Limiting Noise}

\subsection{The Shot Noise on the CCD pixel signal}

Since laser emission and photodetection on a CCD camera  pixel are random processes,  the signal that is obtained on a
CCD pixel exhibits Poisson noise. The effect of this Poisson noise, which cannot be avoided, on the holographic signal
and on the holographic reconstructed images, is the Ultimate Theoretical Limiting noise, which we will study here.

We can split the  signal  $I_{k,p,q}$  we get for  frame $k$ and pixel $(p,q)$ in a noiseless  average component and a
noise component:
\begin{equation}\label{EQ_noise on Ikpq}
    I_{k,p,q} \equiv  \langle I_{k,p,q} \rangle + i_{k,p,q}
\end{equation}
where $\langle ~\rangle$ is the statistical average operator,
and  $i_{k,p,q}$ the noise component. To go further in the
discussion, we will use photo electrons Units to measure the
signal $I_{k,p,q}$.

We must notice that the local oscillator signal ${\cal{E}}_{LO}$
is large, and corresponds to a large number of photo electrons
(e). In real life, this assumption is true. For example,   if we
adjust the power of the LO beam to be at the half maximum of the
camera signal in DC unit (2048 DC in our case), the
pixel signal will be about $10^4$~e, since the ''Camera Gain''
of our camera is $4.8~$e per DC. There are two consequences which
simplify the analysis.\begin{itemize}
    \item {
First, the signal $I_{k,p,q} $ exhibits a gaussian distribution
around its statistical average.}
    \item {
Second, both the quantization noise of  the photo electron
signal ($I_{k,p,q}$ is an integer in photo electron Units), and
the quantization noise of   the Digital Count signal
($I_{k,p,q}$ is an integer in DC Units) can be neglected. These
approximations are valid, since the width of the $I_{k,p,q} $
gaussian distribution is much larger than one in both photo
electron and DC Units. In the example given above, $\langle
I_{k,p,q} \rangle \simeq 10^4$, and this width is $\simeq 10^2$
in photo electron Units, and $\simeq 20$ in DC Units.}
\end{itemize}
One can thus consider that  $I_{k,p,q} $, $\langle I_{k,p,q} \rangle$ and $i_{k,p,q}$ are  floating numbers (and not
integer). Moreover, $i_{k,p,q}$ is a zero-average random Gaussian distribution, with
\begin{equation}\label{Eq_Ikpq variance}
   \langle i_{k,p,q}^2 \rangle = \langle I_{k,p,q} \rangle
\end{equation}

To analyze the LO shot noise contribution to the holographic signal $H_{p,q}$, one of the most simple method is to
perform Monte Carlo simulation from Eq.\ref{EQ_noise on Ikpq} and Eq.\ref{Eq_Ikpq variance}. Since $I_{k,p,q}$ is ever
large (about $10^4$ in our  experiment),  $\langle I_{k,p,q} \rangle $ can be replaced by $ I_{k,p,q} $ (that results
from measurements) in the right member of Eq.\ref{Eq_Ikpq variance}. One has thus:
\begin{equation}\label{Eq_Ikpq variance2}
   \langle i_{k,p,q}^2\rangle = \langle I_{k,p,q} \rangle \simeq I_{k,p,q}
\end{equation}
Monte Carlo simulation of the noise can be done from Eq.\ref{EQ_noise on Ikpq} and Eq.\ref{Eq_Ikpq variance2}
%

\subsection{The Object field Equivalent Noise  for 1 frame}

\begin{figure}[h!]
\begin{center}
\includegraphics[width=8cm,keepaspectratio=true]{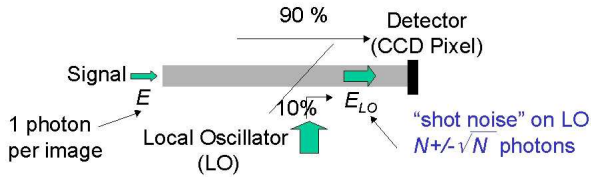}
\caption{ 1 photon equivalent signal (accounting Heterodyne gain),  and shot noise on the holographic Local Oscillator
beam. }\label{fig2_shot_noise}
\end{center}
\end{figure}

In order to discuss the effect of the shot noise on the heterodyne
signal ${\cal E}_{p,q} {\cal E}_{LO}^*$ of Eq.\ref{Eq_slide_2}, let
us consider the simple situation sketched on
Fig.\ref{fig2_shot_noise}. A weak object field $ {{E}}$, with  1
photon or 1 photo electron per pixel and per frame, interferes with a
LO field $ {{E}}_{LO}$ with $N  $ photons, where $N$ is large ($N =
10^4$, in the case of our experiment). Since the LO beam signal
$a^2T|{\cal E}_{LO}|^2$ is equal to $N $ photons, and the object
field signal  $a^2T|{\cal E}_{p,q}|^2$ is one photon, we have:
\begin{equation}\label{Eq_equiv_noise_1}
    I_{k,p,q} = N + 1 + i_{k,p,q} + a^2T{\cal E}_{p,q} {\cal E}_{LO}^* e^{...} + c.c.
\end{equation}
Note that  the heterodyne
signal ${\cal E}_{p,q} {\cal E}_{LO}^*$ is much larger than $|{\cal E}_{p,q}|^2$. This is the
gain effect, associated to the  coherent detection of the field ${\cal E}_{p,q} $. This gain is commonly called
''heterodyne gain'', and is proportional to the amplitude of the LO field ${\cal E}_{LO}^*$.

The purpose of the present discussion is to determine the effect of the noise term $i_{k,p,q}$ of
Eq.\ref{Eq_equiv_noise_1} on the holographic signal $H_{p,q}$. Since $H_{p,q}$ involves only the heterodyne term
${\cal E}_{p,q} {\cal E}_{LO}^*$ (see  Eq.\ref{Sum_heterodyne_2}), we have to compare, in Eq.\ref{Eq_equiv_noise_1},
\begin{itemize}
   \item {the shot noise term $i_{k,p,q}$.}
   \item {and the heterodyne term ${\cal E}_{p,q} {\cal E}_{LO}^*$}
  \end{itemize}

Let's consider first the shot noise term.   We have
\begin{equation}\label{Eq_equiv_noise_3}
    \langle i_{k,p,q}^2 \rangle  = \langle I_{k,p,q}\rangle = N + 1 \simeq N
\end{equation}
The variance of the shot noise term is thus $\sqrt{N}=10^2$. Since this noise is mainly related to the shot noise on
the local oscillator (since $N\gg 1$), one can group together, in Eq.\ref{Eq_equiv_noise_1}, the LO beam term (i.e.
$N$) with the noise term $i_{k,p,q}$, and consider that the LO beam signal fluctuates,  the number of LO beam photons
being thus ''$N \pm \sqrt{N}$'', as mentioned on Fig.\ref{fig2_shot_noise}.

Consider now the the heterodyne beat signal. Since we have $N$ photons on the LO beam, and 1 photon on the object beam,
we get:
\begin{equation}\label{Eq_equiv_noise_2}
    a^2T |{\cal E}_{p,q} {\cal E}_{LO}^*| \equiv \left( \left(a^2T{|\cal E}_{p,q}|^2\right) \left(a^2T{|\cal
    E}_{LO}|^2\right)\right)^{1/2}=N^{1/2}
\end{equation}
The heterodyne beat signal ${\cal E}_{p,q} {\cal E}_{LO}^*$  is thus $\sqrt{N}=10^2$.

The shot noise term  $ i_{k,p,q}$  is thus equal to the heterodyne signal ${\cal E}_{p,q} {\cal E}_{LO}^*$
corresponding to 1 photon on the object field. This means that shot noise $ i_{k,p,q}$ yields  an equivalent noise of 1
photon per pixel, on the object beam. This result is obtained here for 1 frame. We will show that it remains true for a
sequence of $4n$ frames, whatever $4n$ is.

\subsection{The Object field Equivalent Noise  for $4n$ frames}

Let us introduce the  DC component signal $D$, which is similar to the heterodyne signal $H$ given by
Eq.\ref{Sum_heterodyne}, but without phase factors:
\begin{equation}\label{DC_term}
D \equiv \sum_{k=0}^{4n-1} I_k
\end{equation}
The component $D$ can be defined for each pixel $(p,q)$ by :
\begin{equation}\label{DC_term_pq}
D_{p,q} \equiv \sum_{k=0}^{4n-1} I_{k,p,q}
\end{equation}
%
Since $I_{k,p,q}$ is always large in real life (about $10^4$ in our experiment), the shot noise term can be neglected  in
the calculation of $ D_{p,q}$ by  Eq.\ref{DC_term_pq}. We have thus:
\begin{equation}\label{DC_term_pq_1}
D_{p,q} \equiv  \sum_{k=0}^{4n-1} I_{k,p,q}   = 4n a^2T \left(|{\cal E}_{p,q}|^2+|{\cal E}_{LO}|^2\right)
\end{equation}

We are implicitly interested  by the low signal situation (i.e. ${\cal E}_{p,q}\ll {\cal E} _{LO}$ ) because we focus
on noise analysis. In that case, the $|{\cal E}_{p,q}|^2$ term can be neglected in Eq.\ref{DC_term_pq_1}. This means
that $D_{p,q}$ gives a good approximation for the  LO signal.
\begin{equation}\label{DC_term_pq_11}
D_{p,q}  \equiv \sum_{k=0}^{4n-1} I_{k,p,q}   \simeq  4n a^2T |{\cal E}_{LO}|^2
\end{equation}
We can get then the signal field  $|{\cal E}_{p,q}|^2$ from Eq.\ref{Sum_heterodyne_2} and Eq.\ref{DC_term_pq_11}:
\begin{equation}\label{Eq_S_ACoverDC}
 \frac{|H_{p,q}|^2}{D_{p,q}} \simeq 4n a^2 T  |{\cal E}_{p,q}|^2
\end{equation}
In this equation, the ratio ${|H_{p,q}|^2}/{D_{p,q}}$ is
proportional to the number of frames of the sequence ($4n$), This
means that ${|H_{p,q}|^2}/{D_{p,q}}$  represents the signal field
$|{\cal E}_{p,q}|^2$ summed over the all frames.

Let us calculate the effect of the shot noise on
${|H_{p,q}|^2}/{D_{p,q}}$. To calculate this effect, one can make a
Monte Carlo simulation as mentioned above, but a simpler calculation
can be done here. In Eq.\ref{Eq_S_ACoverDC}, we develop $|H_{p,q}|$
in  statistical average and noise components (as done for
$I_{k,p,q}$ in Eq.\ref{EQ_noise on Ikpq}), while neglecting noise in
$D_{p,q}$.

We get:
\begin{eqnarray}\label{Eq_S_ACoverDC_222}
\left \langle \frac{|H_{p,q}|^2}{D_{p,q}} \right \rangle
 \simeq  \frac {1}{ \langle D_{p,q}\rangle }
&\times& ( | \langle H_{p,q} \rangle |^2
    + \langle |h_{p,q}|^2 \rangle \\
\nonumber
    &+& \langle \langle H_{p,q} \rangle h_{p,q}^* \rangle
    + \langle \langle H_{p,q}^* \rangle h_{p,q}
    )
\end{eqnarray}
where
\begin{equation}\label{Eq_nois_on_hpq}
    H_{p,q}= \langle H_{p,q} \rangle  +  h_{p,q}
\end{equation}
with
\begin{equation}\label{Eq_petit_hpq}
    h_{p,q} = \sum_{k=0}^{4n-1} j^k  i_{k,p,q}
\end{equation}
which is the shot noise random contribution to  $H_{p,q}$. In
Eq.\ref{Eq_S_ACoverDC_222}  the $\langle \langle H_{p,q} \rangle
h_{p,q}^* \rangle$ term is zero since  $ h_{p,q}^*$ is random while
$\langle H_{p,q} \rangle$ is not random. The two terms $\langle
\langle H_{p,q} \rangle h_{p,q}^* \rangle$  and $\langle \langle
H_{p,q}^* \rangle h_{p,q} \rangle$ can be thus removed.

On the other hand, we get for $|h_{p,q}|^2$
\begin{equation}\label{Eq_hpq2}
   |h_{p,q}|^2 = \sum_{k=0}^{4n-1}   |i_{k,p,q}|^2 +
   \sum_{k=0}^{4n-1}
   \sum_{k'=0,  k' \ne k}^{4n-1}  j^{k-k'} i_{k,p,q} i_{k',p,q}
\end{equation}
Since $i_{k,p,q}$ and $ i_{k',p,q}$ are uncorrelated,  the
$i_{k,p,q} i_{k',p,q}$ terms cancel in the calculation of the
statistical average of $ |h_{p,q}|^2$. We get then  from
Eq.\ref{Eq_Ikpq variance}
\begin{equation}\label{Eq_hpq2_aver}
  \langle  |h_{p,q}|^2 \rangle = \sum_{k=0}^{4n-1}  \langle |i_{k,p,q}|^2 \rangle
  = \sum_{k=0}^{4n-1} \langle  I_{k,p,q} \rangle = \langle D_{p,q} \rangle
\end{equation}
and Eq.\ref{Eq_S_ACoverDC_222} becomes:
\begin{equation}\label{Eq_S_ACoverDC_222_bis}
   \left \langle \frac{|H_{p,q}|^2}{D_{p,q}} \right \rangle  = \frac { | \langle
H_{p,q} \rangle |^2}{ \langle D_{p,q} \rangle} +1
\end{equation}

This equation means that the average detected  intensity signal $\langle {|H_{p,q}|^2}/{D_{p,q}}\rangle $ is the sum of
the square of the average object field  $\langle |H_{p,q}|\rangle/\sqrt{\langle D_{p,q} \rangle} $ plus one
photo-electron. Without illumination of the object, the average object field is zero, and the detected signal is 1
photo-electron.  The equation establishes thus that the LO shot noise yields a signal intensity corresponding exactly 1
photo-electron per pixel whatever the number of frames $4n$ is.

The 1 e noise floor we get here can be also interpreted as resulting from the heterodyne detection of the vacuum field
fluctuations~\cite{bachor1998guide}.

\subsection{The detection bandwidth, and the noise}

From a practical point of view, the holographic detected signal intensity increases linearly with the acquisition time
$4nT$ (since ${|H_{p,q}|^2}/{D_{p,q}}\propto 4n$), while the noise contribution remains constant: the 1 e noise
calculated by Eq.\ref{Eq_S_ACoverDC_222} corresponds to a sequence of $4n$ frames, whatever the number $4n$ of frames
is. The coherent character of holographic detection explains this paradoxical result.

\begin{figure}[h!]
\begin{center}
\includegraphics[width=7cm,keepaspectratio=true]{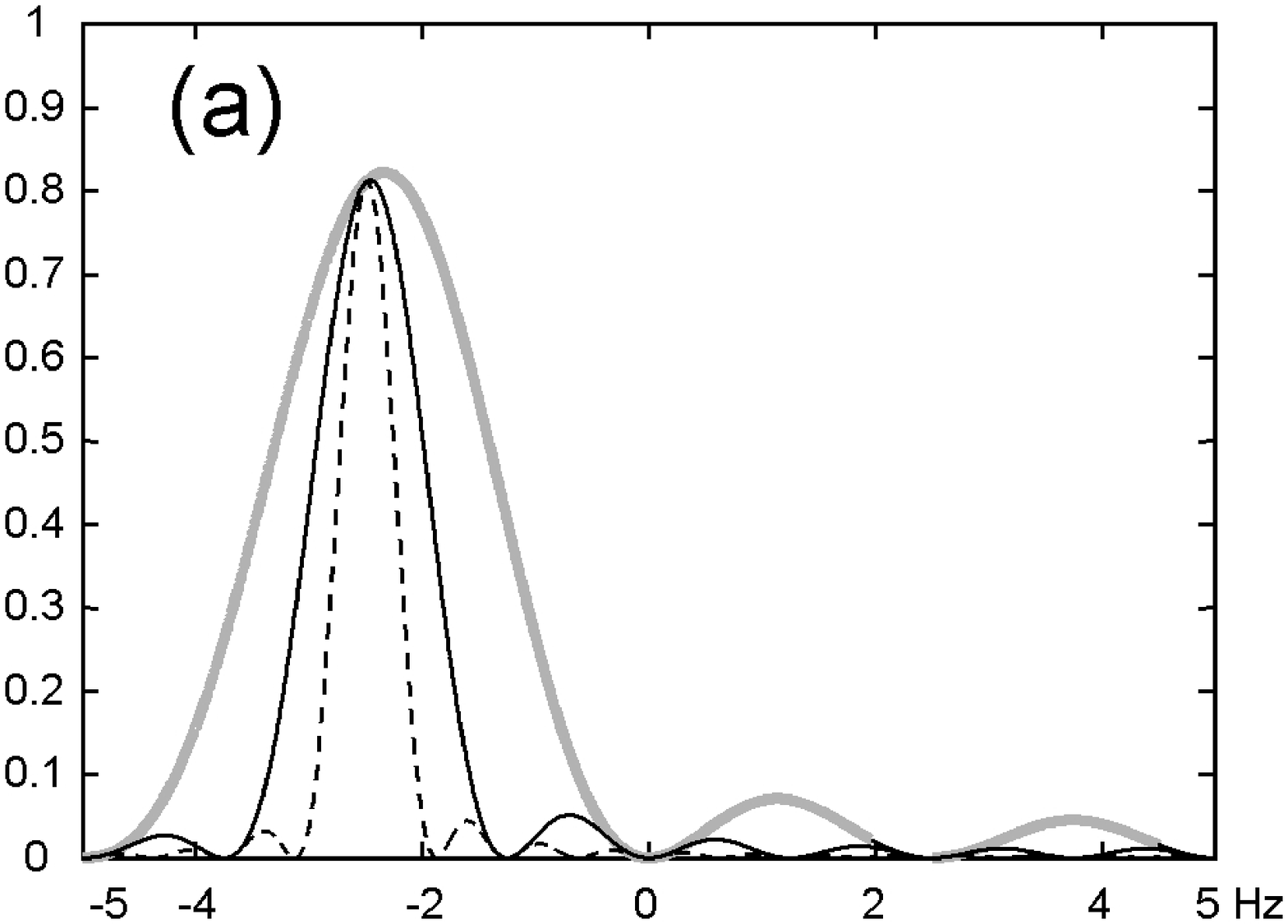}
\includegraphics[width=7cm,keepaspectratio=true]{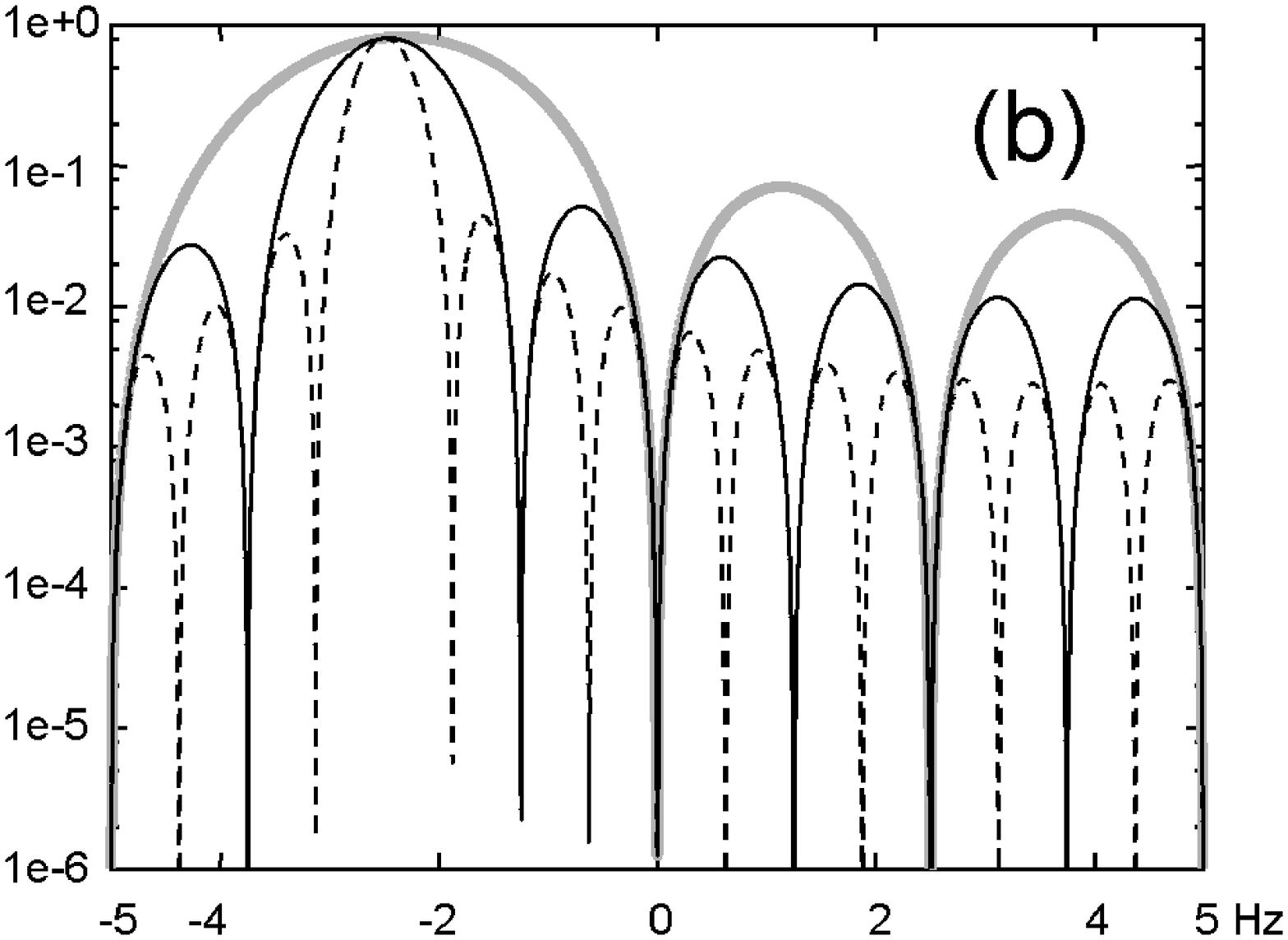}
\caption{Frequency  response $|\eta(x)|^2$ for heterodyne signal in intensity, as a function of the heterodyne beat
frequency $x=f-f_{LO}$ for sequences of $4n$ frames with $4n=4$ (heavy grey line), $4n=8$ (solid black line), and
$4n=16$ (dashed black line). Calculation is done for $T=0.1$~s.  Vertical axis axis is  $|\eta(x)|^2$ in linear (a) and
logarithmic (b) scales.  Horizontal axis is $x=f-f_{LO}$ in Hz.}\label{fig3_heterodyne_freq_response_theo}
\end{center}
\end{figure}

The noise remains constant with time because the noise is broadband (it is a white noise), while the detection is
narrowband. The noise that is detected is proportional to the product of the exposure time, which is proportional to
the acquisition time $4nT$, with the detection Bandwidth, which is inversely proportional to $4nT$. It does not depend
thus on $4nT$.

To illustrate this point, we have calculated, as a function of the exposure time $4nT$, the frequency response of the
coherent detection made by summing the $4n$ frames with the phase factors $j^k$ of Eq.\ref{Sum_heterodyne}. Let us call
$\eta$ the detection efficiency  for the signal field complex amplitude. We get:
\begin{eqnarray}\label{Eq_heterodyne_detection_response}
    \eta(x)&=& \frac{1}{4nT}\sum_{k=0}^{4n-1}j^k\int_{t=kT-T/2}^{kT+T/2}e^{j2\pi xt} dt\\
     &=&\textrm{sinc}(\pi x T) \times \frac{1}{4n}  \sum_{k=0}^{4n-1} j^k  {\rm e}^{j 2 \pi k x T }
\end{eqnarray}
Here $x=f-f_{LO}$ is the heterodyne beat frequency; $f$ is the optical frequency of the signal beam, and
$f_{LO}$ the frequency of the LO beam. In equation \ref{Eq_heterodyne_detection_response}, the factor $\textrm{sinc}(\pi x T)$
corresponds to the integration of the beat signal, whose frequency $x$ is non zero, over the  CCD frame finite exposure
time $T$. The summation over the frames $k$ of Eq.\ref{Sum_heterodyne} yields, in
Eq.\ref{Eq_heterodyne_detection_response},  to sum the phase  $ {\rm e}^{j 2 \pi k  T x}$  of the heterodyne beat at the
beginning of each frame $k$ with the phase factor $j^k$. To the end, the factor $1/4n$ in
Eq.\ref{Eq_heterodyne_detection_response} is a normalization factor that is the inverse of the number of terms within
the summation over $k$. This $1/4n$ factor keeps the maximum of $|\eta(x)|$ slightly lower than 1.

We have calculated,  and plotted on Fig.\ref{fig3_heterodyne_freq_response_theo}, the detection frequency spectrum
$|\eta(x)|^2$ for sequences with different number of frames $4n$. The heavy grey line curve corresponds to 4 frames,
the solid line curve to 8 frames, and the dashed line to 16 frames. As seen, the width of the frequency response
spectrum (and thus the frequency response area) is inversely proportional to the exposure time ($(4\,T)^{-1}$, $(8\,T)^{-1}$ and $(16\,
T)^{-1}$ respectively).

\begin{figure}[h!]
\begin{center}
\includegraphics[width=7cm,keepaspectratio=true]{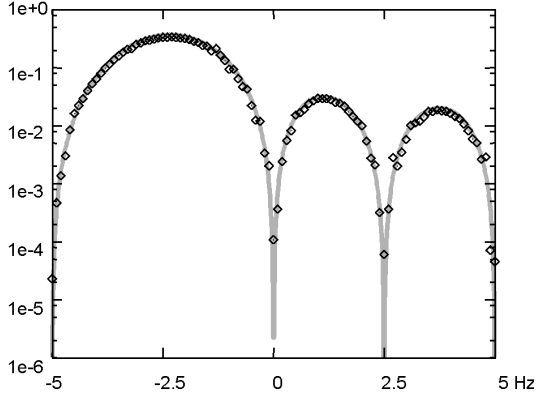}
\caption{Frequency  response  for heterodyne signal in intensity, as a function of the heterodyne beat frequency
$x=f-f_{LO}$ for a sequence of $4$ frames: theory $|\eta(x)|^2$ (heavy grey line), and experiment $W(x)$ (points).
Calculation and experiment are done with $T=0.1$~s.  Vertical axis axis is  $|\eta(x)|^2$ or $W(x)$ in logarithmic
scales. Horizontal axis is $x=f-f_{LO}$ in Hz. }\label{fig4_heterodyne_freq_response_exp}
\end{center}
\end{figure}

To verify the validity of Eq.\ref{Eq_heterodyne_detection_response}, we have swept the frequency $f_{LO} = \omega_{LO}/(2\pi)$ of our
holographic LO by detuning the AOMs frequency (see Fig.\ref{fig1_setup1}), while keeping constant the illumination frequency $f$. We have then measured the weight $W(x)$ of the reconstructed holographic intensity signal
$H^2$ as a function of the beat frequency $x=f-f_{LO}$.
Figure \ref{fig4_heterodyne_freq_response_exp} shows the comparison of the theoretical signal $|\eta(x)|^2$ (heavy grey
line), with the experimental data $W(x)$ (points). The agreement is excellent.

\section{Reaching the theoretical Shot Noise in experiment}

In the previous sections, we have shown that the theoretical noise  on the holographic reconstructed intensity images
is 1 photo electron per pixel whatever the number of recorded frames is. We will now discuss the ability to reach
this limit in real time holographic experiment. Since we consider implicitly a very weak object beam signal, the
noises that must be considered are
\begin{itemize}
    \item {the read noise of the CCD camera,}
   \item {the quantization noise of the camera A/D  converter,}
    \item {the technical noise on the LO beam,}
        \item {and the LO beam shot noise, which yields the theoretical noise limit.}
\end{itemize}

\begin{figure}[h!]
\begin{center}
\begin{tabular}{|c|c|c|}
  \hline
  Number of pixels & 1280 (H) $\times$ 1024 (V) \\
  Pixel size & $6.7 \times 6.7 \mu$m \\
  Frame Rate & 12.5~fps \\
 Full Well Capacity  & 25~000 e \\
  A/D Converter  & 12 bits: 0... 4095 DC \\
  A/D conversion factor (Gain) & 4.8 e/DC \\
  QE $@$~500~nm :  & 40 \% \\
  QE $@$~850~nm :  & 6\%   \\
  Read Noise  & 20  e \\
  Dark Noise  & 3 e/sec/pix \\
   \hline
\end{tabular}
\caption{Main characteristics of the PCO Pixelfly Camera} \label{fig5_camera_datasheet}
\end{center}
\end{figure}

\subsection{The technical noise within the $(k_x,k_y)$ reciprocal
space}

The main characteristics of our camera are given in Fig.\ref{fig5_camera_datasheet}. In a typical experiment, the LO
beam power is adjusted in order to get $2000$~DC on the A/D Converter, i.e. about $10^4$~e on the each CCD pixel. The
LO shot noise, which  is about 100~e, thus much larger than the Read Noise (20~e), than the Dark Noise (3~e/sec), and
than the A/D converter quantization noise (4.8~e, since 1 DC corresponds to 4.8~e). The noise of the camera, which can
be neglected, is thus not limiting in reaching the noise theoretical limit.

The LO beam  that reaches the camera is essentially flat field
(i.e. the field intensity $|{\cal E}_{LO}|^2$ is the same for
all the pixels). The LO beam technical noise is thus highly
correlated from pixel to pixel. This is for example the case of the
noise induced by the fluctuations of the main laser  intensity,
or by the vibrations of the mirrors within the LO beam arm. To
illustrate this point, we have recorded a sequence  of $4n=4$
frames $I_k$  with LO beam, but  without signal (i.e. without
illumination of the object). We have recorded thus the hologram of
the ''vacuum field''. We have calculated then the complex
hologram $H$ by Eq.\ref{Sum_heterodyne}, and the reciprocal
space hologram $\tilde H$ by FFT (i.e. by Eq.\ref{eq_FFT1}).

\begin{figure}[h!]
\begin{center}
\includegraphics[width=7cm,keepaspectratio=true]{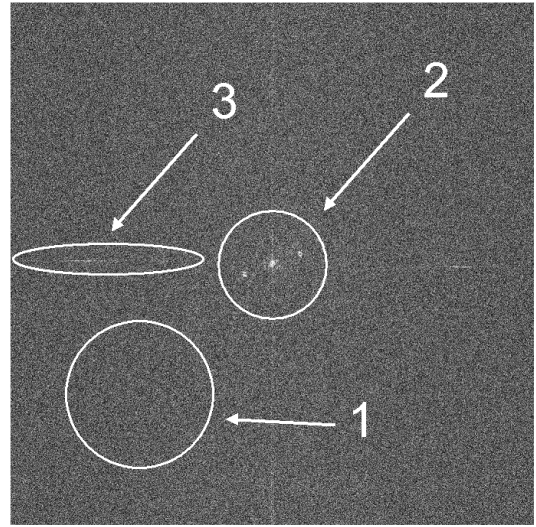}
\caption{
Intensity image of $\tilde{H}(k_x,k_y,0)$  for $4n=4$ frames without signal $\cal E$. Three kind
of noises can be identified. left : FFT aliasing, down left : shot noise, middle : technical noise of the CCD. By
truncating the image and keeping only the left down part, the shot noise limit is reached.
The image is displayed in arbitrary logarithm grey scale.
 }\label{fig6_image_shot_noise_zero_signal_exp}
\end{center}
\end{figure}

The reciprocal space holographic  intensity $|\tilde H|^2$  is
displayed on Fig.\ref{fig6_image_shot_noise_zero_signal_exp} in arbitrary logarithm grey
scale.  On most of the reciprocal space (within for example
circle 1),  $|\tilde H|^2$ corresponds to a random speckle whose
average intensity is uniformly distributed along $k_x$ and
$k_y$. One observes nevertheless bright points within circle 2,
which corresponds to $(k_x,k_y) \simeq (0,0)$. These points
correspond to the technical noise, which is flat field within
the CCD plane $(x,y)$, and which has thus a low spatial
frequency spread within the $(k_x,ky)$ reciprocal space. One
see also, on the Fig.\ref{fig6_image_shot_noise_zero_signal_exp} image, an horizontal and a
vertical bright line, which  corresponds to $k_y=0$ and $k_x=0$
(zone 3 on Fig.\ref{fig6_image_shot_noise_zero_signal_exp}). These parasitic bright lines
are related  to Fast Fourier Transform aliases, that are related
to the discontinuity of the signal $I_k$ and $H$ at edge of the
calculation grid, in the $(x,y)$ space.

We have  measured $\langle| {\tilde H}|^2 \rangle$ by replacing
the statistical average $\langle~\rangle$ by a spatial average
over a region of the conjugate space without technical noise
(i.e. over region 1). This gives a measurement of $\langle|
{\tilde H}|^2 \rangle$, i.e. a measurement of $\langle |  H|^2
\rangle$, since the space average of $| {\tilde H}|^2$ and $| {
H}|^2$ are equal, because of the FT Parceval theorem. We have
also measured $D$ from the sequence of frames $I_k$ (see
Eq.\ref{DC_term}). Knowing the A/D conversion factor (4.8 e/DC),
we have calculated the noise intensity  $\langle| {\tilde H}|^2
\rangle/ \langle D \rangle$ in photo-electron units, and we get,
within 10\%, $1$ photo electron per pixel, as expected
theoretically for the shot noise (see
Eq.\ref{Eq_S_ACoverDC_222}).

This result proves that it is possible to perform shot noise
limited holography in actual experiments. Since the low spatial frequency region of the reciprocal space (region 2) must be avoided because of
the technical noise, it is necessary  to perform digital holography
in an off-axis configuration, in order to reach the
Eq.\ref{Eq_S_ACoverDC_222} shot noise limit.

\subsection{Effect the finite size of the pixel.}

\begin{figure}[h!]
\begin{center}
\includegraphics[width=7cm,keepaspectratio=true]{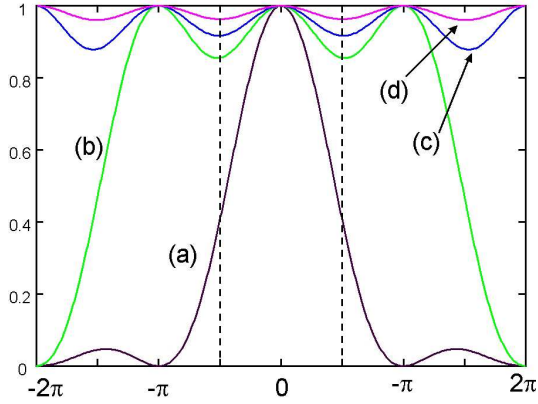}
\caption{One dimension (1D) angular response of the detection efficiency for the intensity $\sum_m
|\textrm{sinc}(X+m\pi)|^2$ as a function of $X$ for the main lobe: $m=0$ (a); for the main  lobe and 2 aliases: $m=0,
\pm 1$ (b);  for the main lobe and 4 aliases: $m=0, \pm 1, \pm 2$ (c); for the main  lobe and 10 aliases:  $m=0, \pm 1,
\pm 2,...\pm 5$ (d).
 }\label{fig_fig_ang_response}
\end{center}
\end{figure}

Because of the finite size  of the pixels $d_{pix}$, the heterodyne detection efficiency within direction $k_x,k_y$ is
weighted by a factor $\zeta$ for the field $\tilde H$, and $|\zeta|^2$ for the intensity $|\tilde H|^2$ with:
\begin{eqnarray}\label{Eq_eta2}
\nonumber    \zeta(k_x,k_y) &=&\frac{1}{d_{pix}^{~2}}~\int_{x=\frac{-1}{2}d_{pix}}^{\frac{1}{2}d_{pix}}
 \int_{y=\frac{-1}{2}d_{pix}}^{\frac{1}{2}d_{pix}} e^{j(k_x x + k_y y)}dx~dy\\
&=& \textrm{sinc}(X)  \textrm{sinc}(Y)
\end{eqnarray}
with $X=k_x d_{pix}/2$ and $Y=k_y d_{pix}/2$. This factor $\zeta$ corresponds to the angular $\textrm{sinc}$
diffraction pattern of the rectangular pixels, which affects the component of $\tilde H$ corresponding to the signal of
the object. The efficiency in energy $|\zeta|^2$ is plotted in Fig.\ref{fig_fig_ang_response}, curve  (a) in black.

Because of the sampling made by the  CCD pixels, the hologram $\tilde H(k_x,k_y)$ is periodic in the reciprocal space,
with a periodicity equal to $2\pi/d_{pix}$ for $k_x$ and $k_y$, or $\pi$ for $X$ and $Y$.  This means that the edges of
the FFT calculation grid, which are displayed on Fig.\ref{fig_fig_ang_response} as vertical dashed lines, corresponds
to $k_x,k_y=\pm \pi/d_{pix}$ or to $X,Y=\pm \pi/2$. Note that the detection efficiency is non zero at the edges of the
calculation  grid since we have $|\zeta|^2 =4/\pi^2 \simeq 0.40$  for $X=\pi/2$ and $Y=0$.

If the factor $|\zeta|^2$ affects the component of $|\tilde H|^2$ corresponding to the signal of the object, it do not
affects the shot noise component, whose weight is 1 whatever $k_x$ and $k_y$ are. One can demonstrate this result by
calculating the noise by Monte Carlo simulation  from Eq.\ref{EQ_noise on Ikpq} and Eq.\ref{Eq_Ikpq variance2}. The
Monte Carlo simulation yields a fully random speckle noise, both in the $x,y$ space, and in the $k_x,k_y$ reciprocal
space.

This point can be understood another way, which is illustrated by Fig.\ref{fig_fig_ang_response}. Each pixel is a
coherent detector, whose detection antenna diagram is the Fig.\ref{fig_fig_ang_response} (a) $\textrm{sinc}$ function.
Because of the periodicity within the reciprocal space, the signal that is detected for $(k_x,k_y)$ or for $(X,Y)$
corresponds to the sum of the signal within the main lobe $(X,Y)$, and within all the aliases corresponding to the
periodicity $(X+m\pi,Y+m'\pi)$. Since the object is located within a well defined direction, the main lobe contribute
nearly alone for the signal. But this is not true for the shot noise, since the shot noise (or the vacuum field noise)
spreads over all $(k_x,k_y)$ points of the reciprocal space with a flat average density. One has thus to sum the
response of the main lobe (i.e. $|\textrm{sinc}(X)|^2$ in 1D) with all the periodicity aliases (i.e.
$|\textrm{sinc}(X+m\pi)|^2$ with $m\ne 0$). Fig.\ref{fig_fig_ang_response} shows the 1D angular response $\sum_m
|\textrm{sinc}(X+m\pi)|^2$ that correspond to sum of the main lobe  with  more and more aliases. As seen, adding more
and more aliases make the angular response flat and equal to one.

\subsection{Experimental validation with an USAF target.}

\begin{figure}[]
\begin{center}
\includegraphics[width =8.2 cm,keepaspectratio=true]{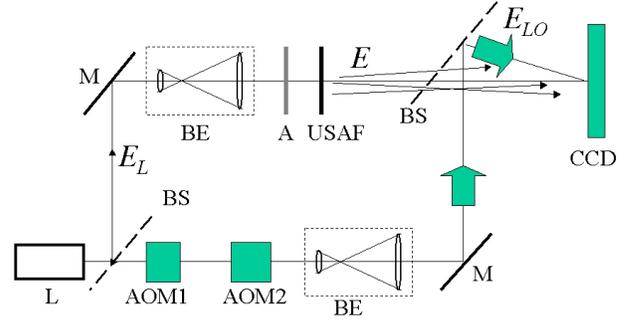}
\caption{ Setup of the test experiment with USAF target. L: main laser; BS: Beam splitter;
AOM1 and AOM2: acousto optic modulators; BE: beam expander;  M:
mirror; A1 and A2: light attenuators. USAF: transmission USAF
target that is imaged. CCD : CCD camera. } \label{fig_setup_usaf}
\end{center}
\end{figure}

We have verified that it is  possible to perform shot noise limited
holography in actual experiments, by  recording the hologram of an USAF
target in transmission. The holographic setup is sketched on
Fig.\ref{fig_setup_usaf}. We have recorded sequences of $4n=12$
frames, and we have reconstructed the image of the USAF target.

\begin{figure}[h!]
\begin{center}
\includegraphics[width=8.5cm,keepaspectratio=true]{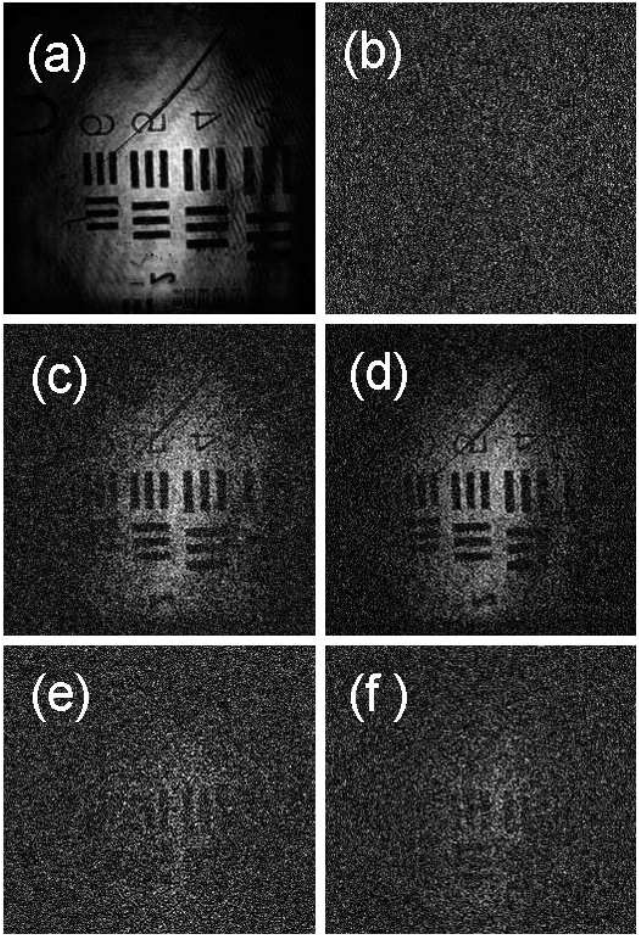}
\caption{(a,c,d): Reconstructions of an USAF target with different level of illumination 700 (a),  1 (c) and 0.15
e/pixel (d). (b): Simulated Shot Noise noise image. (e,f): Simulated reconstructed image obtained by mixing  image (a)
with  weight $X$, and image (b) with weight $1-X$. The weight $X$ is $1/700$ (e), and  $0.15/700$ (f). Images are
displayed in arbitrary logarithmic grey scale.} \label{fig7_usaf_shot}
\end{center}
\end{figure}


Figure \ref{fig7_usaf_shot} shows the holographic reconstructed images of the USAF target.  The intensity of the signal
illumination is adjusted with neutral density filters. In order to filter off the technical noise, the reconstruction
is done by selecting the order 1 image of the object, within the reciprocal space \cite{Cuche00}. Since the  $400\times
400$ pixels region that is selected is off axis, the low spatial frequency noisy region, which corresponds to the zero
order image (region 1 on Fig.\ref{fig6_image_shot_noise_zero_signal_exp}), is filtered-off.

Figure \ref{fig7_usaf_shot} (a,c,d) shows the reconstructed images obtained for different USAF target illumination
levels. For each image, we have  measured
%
%
the average number of photo electrons per pixel corresponding to the object beam, within the reciprocal space region
that has been  selected for the reconstruction (i.e. $400\times 400$ pixels). The  images of Fig. \ref{fig7_usaf_shot}
correspond to 700 (a), 1 (c), and 0.15 e/pix (d) respectively. The  object beam intensity has been measured by the
following way. We have first calibrated the response of our camera with an attenuated laser whose power is known. We
have then measured  with the camera, at high level of signal, the intensity of the signal beam alone (without LO beam).
We have decreased, to the end, the signal beam intensity by using calibrated attenuator in order to reach the low signal
level of the images of Fig. \ref{fig6_image_shot_noise_zero_signal_exp} (a,c,d). In the case of image (a) with
700e/pix, we also have measured the averaged signal intensity from the data by calculating $|H|^2/D$ (see
Eq.\ref{Eq_S_ACoverDC}). The two measurements gave the same result: 700e per pixel.

On figure \ref{fig7_usaf_shot} (a), with 700e per pixel, the USAF signal is much larger than the  shot noise, and the
Sinal to Noise Ratio (SNR) is large. On figure \ref{fig7_usaf_shot} (c), with 1 e per pixel,  the USAF signal is
roughly equal to the shot noise, and the SNR is about 1. With 0.15e per pixel, the SNR is low on
Fig.\ref{fig7_usaf_shot} (d) (about 0.15), and the USAF is hardly seen. It is nevertheless quite difficult to evaluate
the SNR of an image. To perform a more quantitative analysis of the noise within  the images, we have synthesized noisy
images of \ref{fig7_usaf_shot} (e,f) by adding noise to the Fig. \ref{fig7_usaf_shot} (a) noiseless image. We have
first synthesized a pure Shot Noise image , which corresponds to the image that is expected without signal.

The Shot Noise, which is displayed on Fig.\ref{fig7_usaf_shot} (b), is obtained by the following way.  From one of the
measured frames (for example $I_0$) we have calculated the noise components $i_{k,p,q}$ by Monte Carlo drawing with the
condition:
 \begin{equation}\label{Eq_i_k,p,q_I_0,p,q}
    \langle i_{k,p,q}^2 \rangle = I_{0,p,q}
\end{equation}
This condition corresponds to Eq.\ref{Eq_Ikpq variance} since $\langle I_{k,p,q} \rangle \simeq I_{0,p,q}$. We have
then synthesize the sequence of image $I_k$ by:
 \begin{equation}\label{Eq_i_k,p,q synthetise}
    I_{k,p,q} = I_{0,p,q} + i_{k,p,q}
\end{equation}
The Shot Noise image of Fig.\ref{fig7_usaf_shot} (b) is reconstructed then from the $I_{k,p,q}$ sequence.

\begin{figure}[h!]
\begin{center}
\begin{tabular}{|c|c|c|}
  \hline
  Image  & Signal (e/pix)& Noise (e/pix)\\
  \hline
 a & $700$ & 1\\
   b & 0 & 1\\
 c & $ 1$ & 1\\
   d & $ 0.15$ & 1   \\
e & $1$ & 1 \\
 f & $ 0.15$& 1\\
   \hline
\end{tabular}
\caption{Signal and shot Noise on Images of Fig.\ref{fig7_usaf_shot} } \label{fig8_Table 2}
\end{center}
\end{figure}

We have synthesized noisy images by summing the noiseless image of  Fig.\ref{fig7_usaf_shot} (a) with weight $X$, with
the Shot Noise image of Fig.\ref{fig7_usaf_shot} (b) with weight $(1-X)$.   The image of Fig.  \ref{fig7_usaf_shot} (e)
is obtained with $X=1/700$. As shown on the table of Fig.\ref{fig8_Table 2}, Fig.  \ref{fig7_usaf_shot} (e) corresponds
to the same signal, and the same noise than  Figure \ref{fig7_usaf_shot} (c) (1e of signal, and 1e of noise
respectively). Fig. \ref{fig7_usaf_shot} (c) and Fig.  \ref{fig7_usaf_shot} (e), which have been displayed here with
the same linear grey scale,  are visually very similar and exhibit the same SNR. The image of Fig. \ref{fig7_usaf_shot}
is similarly obtained with $X=0.15/700$. It corresponds to the same Signal and Noise  than Figure \ref{fig7_usaf_shot}
(d) (0.15e of signal, and 1e of noise), and, as expected, Fig. \ref{fig7_usaf_shot} (d) and Fig. \ref{fig7_usaf_shot}
(f), which have been displayed here with the same linear grey scale, are similar and exhibit the same SNR too.

Here we demonstrated  our ability to synthesize a noisy image with a noise that is calculated by Monte Carlo from
Eq.\ref{Eq_i_k,p,q_I_0,p,q} and \ref{Eq_i_k,p,q synthetise}. Moreover, we have verified that the noisy image is visually
equivalent to the image we have obtained in experiments. These results  prove that we are able to quantitatively account
theoretically the noise, and that the noise that is obtained in experiments reaches the theoretical limit.

\section{Conclusion}

In this paper we have studied the noise limits in digital holography. We have shown that in high heterodyne gain
of the holographic detection (achieved when the object field power is much weaker than the LO field power), the noise of the CCD camera can be neglected. Moreover by a proper arrangement of the
holographic setup, that combines off-axis geometry with phase shifting acquisition of holograms, it is possible to
reach the theoretical shot noise limit. We have studied theoretically this limit, and we have shown that it corresponds
to 1 photo electron per pixel for the whole sequence of frame that is used to reconstruct the holographic image. This
paradoxical result is related to the heterodyne detection, where the detection bandwidth is inversely proportional to the
measurement time. We have verified all our results experimentally,  and we have shown that is possible to image an object
at very low illumination levels.  We have also shown that is possible to mimic the very weak illumination levels holograms obtained in experiments by Monte Carlo noise modeling. This opens the way to simulation of ''gedanken''  holographic experiments in weak signal conditions.

\section*{Acknowledgment}

The authors would like to thank the ANR (ANR-05-NANO\_031: ''3D NanoBioCell''  grant), and C'Nano Ile de France
(''HoloHetero'' grant).

%


%

\begin{IEEEbiography}[{\includegraphics[width=1in,height=1.25in,clip,keepaspectratio]{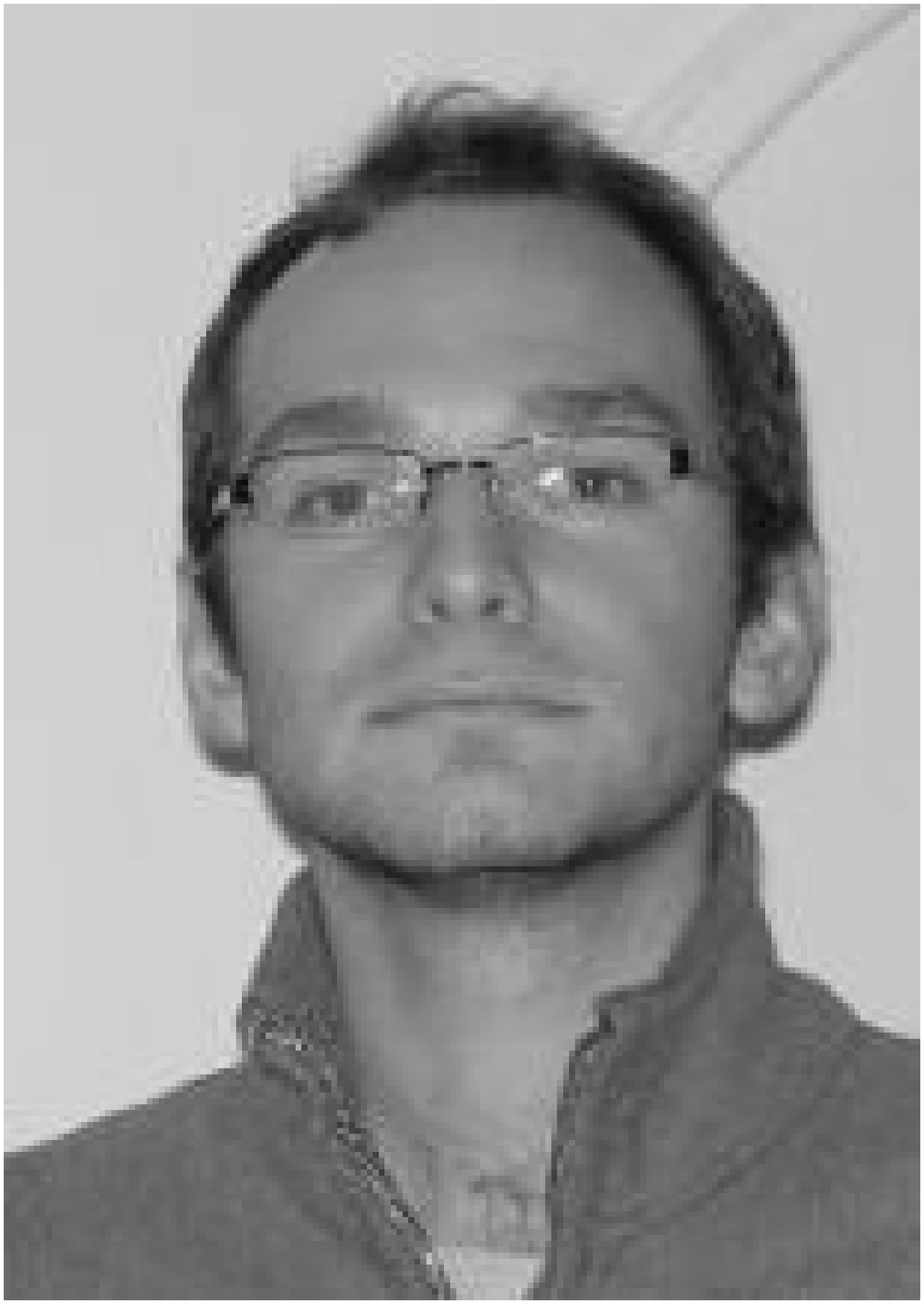}}]{Frédéric~Verpillat}
Frédéric Verpillat entered the Ecole Normale Supérieure of Lyon (France) in 2005, then obtained the M.S degree of the
Ecole Polytechnique Fédérale of Lausanne (Switzerland) in  microengineering in 2009. His specialization is applied
optics for biology or medicine (microscopy, optical tomography). He's preparing a PhD degree at the Laboratoire Kastler
Brossel under the direction of Dr. Michel Gross on the tracking of Nanoparticles with digital holography.
\end{IEEEbiography}

\begin{IEEEbiography}[{\includegraphics[width=1in,height=1.25in,clip,keepaspectratio]{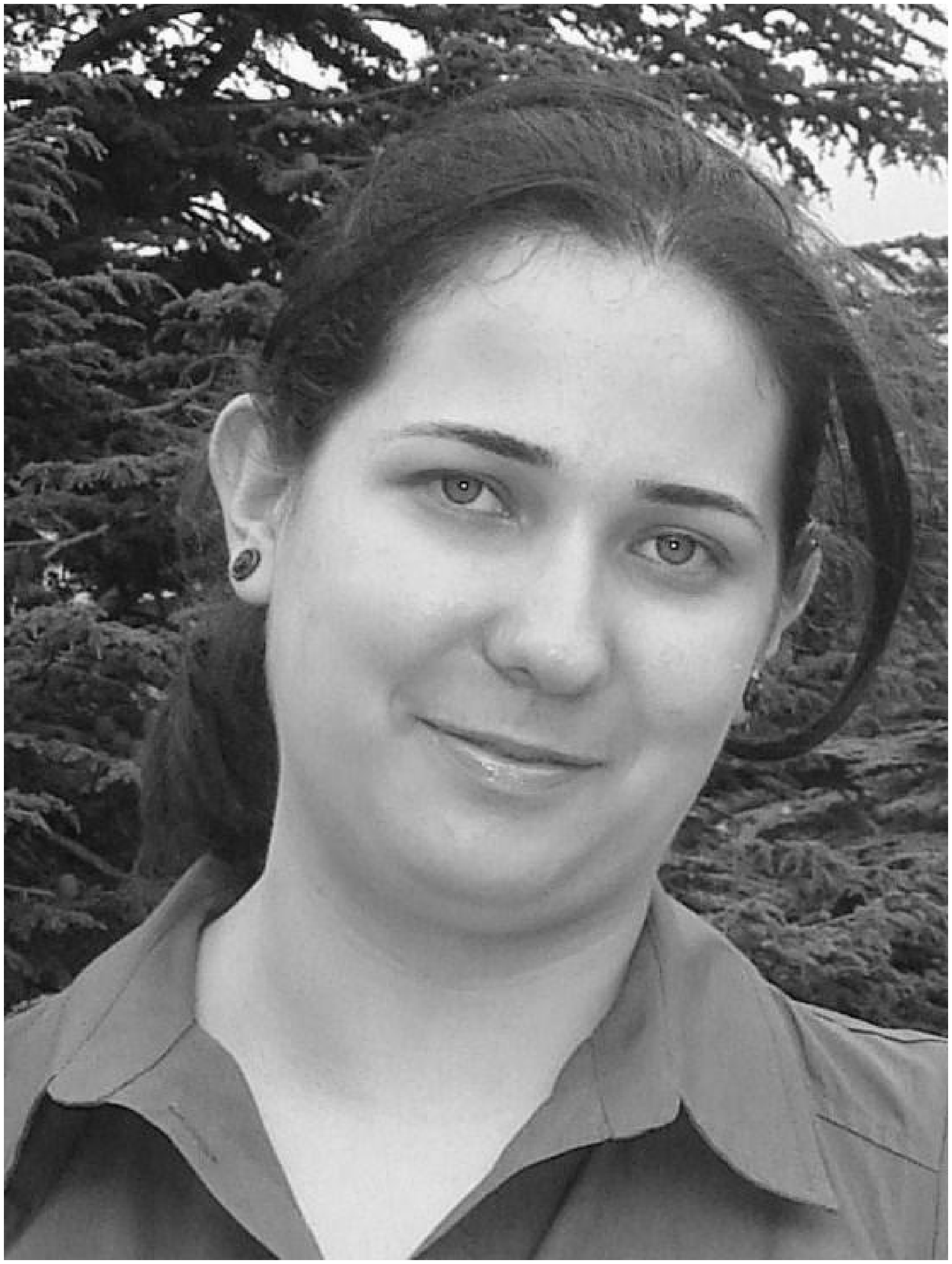}}]{Fadwa~Joud}
Fadwa Joud holds an M.S degree in Condensed Matter and Radiation Physics from Université Joseph Fourrier Grenoble 1,
France. Passionate with NanoBiophotonics, she joined, in October 2008, the team Optics and Nano Objects at the
Laboratoire Kastler Brossel of the École Normale Supérieure in Paris - France to prepare a PhD degree in applied
physics under the supervision of Dr. Michel Gross. Her major research project is Holographic Microscopy and its
applications in the field of Biology and the detection of Nanoparticles.
\end{IEEEbiography}

\begin{IEEEbiography}[{\includegraphics[width=1in,height=1.25in,clip,keepaspectratio]{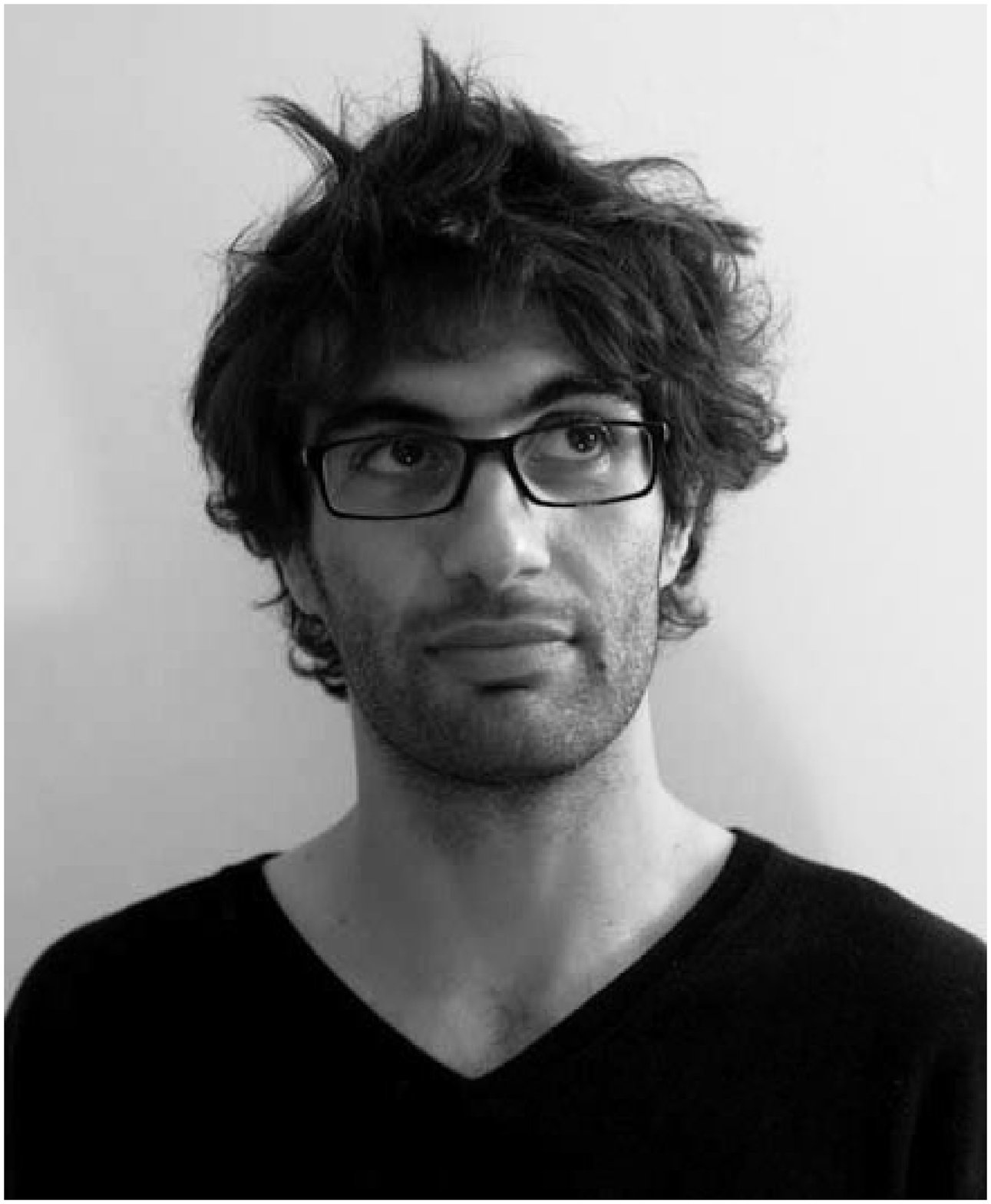}}]{Michael~Atlan}
Michael Atlan is a research investigator at CNRS. He studied to deepen expertise in optical physics during his PhD and
postdocs under the tutelage of Drs. Claude Boccara, Andrew Dunn, Maite Coppey and Michel Gross. He works on
non-invasive and non-ionizing imaging modalities to assess biological structures and dynamic processes from subcellular
to tissular scales, designing coherent optical detection schemes to enable highly sensitive imaging at high throughput.
\end{IEEEbiography}

\begin{IEEEbiography}[{\includegraphics[width=1in,height=1.25in,clip,keepaspectratio]{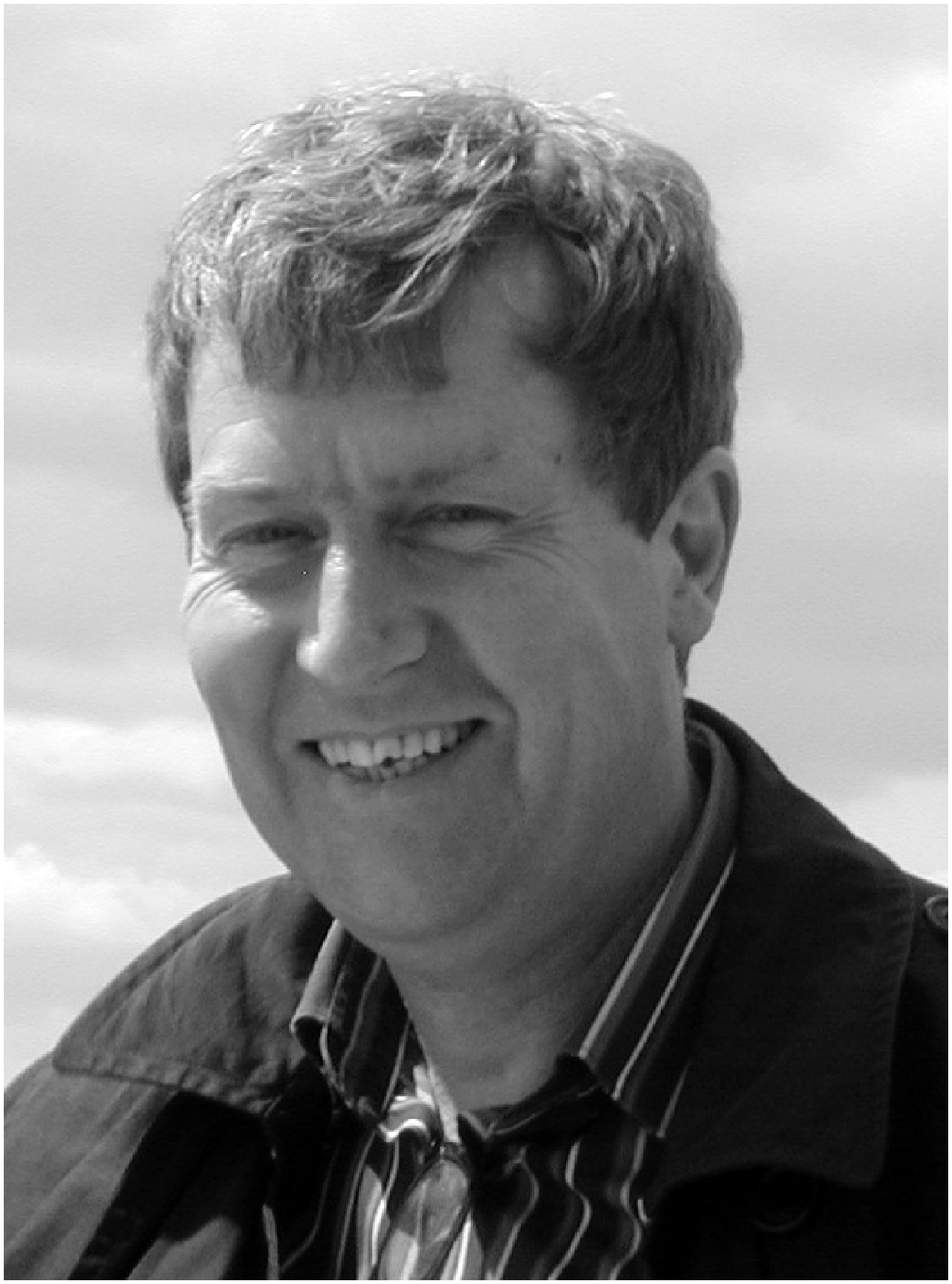}}]{Michel~Gross}
Michel Gross enter the French Ecole Normale Supérieure in 1971. He joined the Laboratoire Kastler Brossel (Paris) in
1975, where he is research scientist. He receive his Ph.D from University Pierre et Marie Curie, Paris, France in 1980.
His scientific interests  are atomic physics (superradiance, Rydberg and  circular atoms), excimer laser refractive
surgery,  millimeter wave and teraherz technology, and digital. He developed a Millimeter Wave Network Analyzer and
participate to the creation the  AB Millimeter company. His main current interest is digital holography. He has
published about 80 scientific papers, and is co-inventor of 6 patents.
\end{IEEEbiography}

\end{document}